\documentclass[a4paper,fleqn,usenatbib]{mnras}

\usepackage[T1]{fontenc}
\usepackage{ae,aecompl}

\usepackage{times}
\usepackage{epsfig}
\usepackage[]{graphicx} 
\usepackage{pdfpages}
\usepackage{verbatim} 
\usepackage[fleqn]{amsmath}
\usepackage{amssymb}
\usepackage{calc}
\usepackage{caption}
\usepackage[normal]{threeparttable}
\usepackage{longtable}
\graphicspath{{./plots/}}
\title[The mean spin temperature from 21\,cm surveys]{Using 21\,cm
  absorption surveys to measure the average \mbox{H\,{\sc i}} spin
  temperature in distant galaxies}

\author[J.~R. Allison et al.]{
J.~R. Allison,$^{1}$\thanks{E-mail: james.allison@csiro.au}
M.~A. Zwaan,$^{2}$
S.~W. Duchesne,$^{3}$ 
and S.~J. Curran$^{3}$
\\
$^{1}$CSIRO Astronomy \& Space Science, PO Box 76, Epping NSW 1710, Australia\\
$^{2}$European Southern Observatory, Karl-Schwarzschild-Str. 2, D85748 Garching, Germany\\
$^{3}$School of Chemical and Physical Sciences, Victoria University of
Wellington, PO Box 600, Wellington 6140, New Zealand}

\date{Accepted XXX. Received YYY; in original form ZZZ}

\pubyear{2016}

\begin{document}
\label{firstpage}
\pagerange{\pageref{firstpage}--\pageref{lastpage}}
\maketitle

\begin{abstract}
  We present a statistical method for measuring the average
  \mbox{H\,{\sc i}} spin temperature in distant galaxies using the
  expected detection yields from future wide-field 21\,cm absorption
  surveys. As a demonstrative case study we consider a simulated
  all-southern-sky survey of 2-h per pointing with the Australian
  Square Kilometre Array Pathfinder for intervening \mbox{H\,{\sc i}}
  absorbers at intermediate cosmological redshifts between $z = 0.4$
  and 1. For example, if such a survey yielded $1000$ absorbers we
  would infer a harmonic-mean spin temperature of
  $\overline{T}_\mathrm{spin} \sim 100$\,K for the population of
  damped Lyman $\alpha$ (DLAs) absorbers at these redshifts,
  indicating that more than $50$\,per\,cent of the neutral gas in
  these systems is in a cold neutral medium (CNM). Conversely, a lower
  yield of only 100 detections would imply $\overline{T}_\mathrm{spin}
  \sim 1000$\,K and a CNM fraction less than $10$\,per\,cent. We
  propose that this method can be used to provide independent
  verification of the spin temperature evolution reported in recent 21
  cm surveys of known DLAs at high redshift and for measuring the spin
  temperature at intermediate redshifts below $z \approx 1.7$, where
  the Lyman-$\alpha$ line is inaccessible using ground-based
  observatories. Increasingly more sensitive and larger surveys with
  the Square Kilometre Array should provide stronger statistical
  constraints on the average spin temperature. However, these will
  ultimately be limited by the accuracy to which we can determine the
  \mbox{H\,{\sc i}} column density frequency distribution, the
  covering factor and the redshift distribution of the background
  radio source population.
\end{abstract}




\section{Introduction}\label{intro}

Gas has a fundamental role in shaping the evolution of galaxies,
through its accretion on to massive haloes, cooling and subsequent
fuelling of star formation, to the triggering of extreme luminous
activity around super massive black holes. Determining how the
physical state of gas in galaxies changes as a function of redshift is
therefore crucial to understanding how these processes evolve over
cosmological time. The standard model of the gaseous interstellar
medium (ISM) in galaxies comprises a thermally bistable medium
(\citealt*{Field:1969}) of dense ($n \sim 100$\,cm$^{-3}$) cold
neutral medium (CNM) structures, with kinetic temperatures of $T_{\rm
  k} \sim 100$\,K, embedded within a lower-density ($n \sim
1$\,cm$^{-3}$) warm neutral medium (WNM) with $T_{\rm k} \sim
10^{4}$\,K. The WNM shields the cold gas and is in turn ionized by
background cosmic rays and soft X-rays (e.g. \citealt{Wolfire:1995,
  Wolfire:2003}). A further hot ($T_{\rm k} \sim 10^{6}$\,K) ionized
component was introduced into the model by \cite{McKee:1977}, to
account for heating by supernova-driven shocks within the inter-cloud
medium. In the local Universe, this paradigm has successfully
withstood decades of observational scrutiny, although there is some
evidence (e.g. \citealt{Heiles:2003b}; \citealt*{Roy:2013b};
\citealt{Murray:2015}) that a significant fraction of the WNM may
exist at temperatures lower than expected for global conditions of
stability, requiring additional dynamical processes to maintain local
thermodynamic equilibrium.

Since atomic hydrogen (\mbox{H\,{\sc i}}) is one of the most abundant
components of the neutral ISM and readily detectable through either
the 21\,cm or Lyman $\alpha$ lines, it is often used as a tracer of
the large-scale distribution and physical state of neutral gas in
galaxies. The 21\,cm line has successfully been employed in surveying
the neutral ISM in the Milky Way
(e.g. \citealt{McClure-Griffiths:2009,Murray:2015}), the Local Group
(e.g. \citealt{Kim:2003,Bruns:2005,Braun:2009,Gratier:2010}) and
low-redshift Universe (see \citealt{Giovanelli:2016} for a
review). However, beyond $z \sim 0.4$ (\citealt{Fernandez:2016})
\mbox{H\,{\sc i}} emission from individual galaxies becomes too faint
to be detectable by current 21\,cm surveys and so we must rely on
absorption against suitably bright background radio (21\,cm) or UV
(Lyman-$\alpha$) continuum sources to probe the cosmological evolution
of \mbox{H\,{\sc i}}. The bulk of neutral gas is contained in
high-column-density damped Lyman-$\alpha$ absorbers (DLAs, $N_{\rm HI}
\geq 2 \times 10^{20}$\,cm$^{-2}$; see \citealt*{Wolfe:2005} for a
review), which at $z \gtrsim 1.7$ are detectable in the optical
spectra of quasars. Studies of DLAs provide evidence that the atomic
gas in the distant Universe appears to be consistent with a
multi-phase neutral ISM similar to that seen in the Local Group
(e.g. \citealt*{Lane:2000}; \citealt*{Kanekar:2001c};
\citealt*{Wolfe:2003b}). However, there is some variation in the cold
and warm fractions measured throughout the DLA population
(e.g. \citealt*{Howk:2005}; \citealt{Srianand:2005, Lehner:2008};
\citealt*{Jorgenson:2010}; \citealt{Carswell:2011, Carswell:2012,
  Kanekar:2014a}; \citealt*{Cooke:2015}; \citealt*{Neeleman:2015}).

The 21-cm spin temperature affords us an important line-of-enquiry in
unraveling the physical state of high-redshift atomic gas. This
quantity is sensitive to the processes that excite the ground-state of
\mbox{H\,{\sc i}} in the ISM
(\citealt{Purcell:1956,Field:1958,Field:1959b,Bahcall:1969}) and
therefore dictates the detectability of the 21\,cm line in absorption.
In the CNM the spin temperature is governed by collisional excitation
and so is driven to the kinetic temperature, while the lower densities
in the WNM mean that the 21\,cm transition is not thermalized by
collisions between the hydrogen atoms, and so photo-excitation by the
background Ly $\alpha$ radiation field becomes important. Consequently
the spin temperature in the WNM is lower than the kinetic temperature,
in the range $\sim$1000 -- 5000\,K depending on the column density and
number of multi-phase components (\citealt{Liszt:2001}). Importantly,
the spin temperature measured from a single detection of extragalactic
absorption is equal to the harmonic mean of the spin temperature in
individual gas components, weighted by their column densities, thereby
providing a method of inferring the CNM fraction in high-redshift
systems.

Surveys for 21\,cm absorption in known redshifted DLAs have been used
to simultaneously measure the column density and spin temperature of
\mbox{H\,{\sc i}} (see \citealt{Kanekar:2014a} and references
therein). There is some evidence for an increase (at $4\,\sigma$
significance) in the spin temperature of DLAs at redshifts above $z =
2.4$, and a difference (at $6\,\sigma$ significance) between the
distribution of spin temperatures in DLAs and the Milky Way
(\citealt{Kanekar:2014a}). The implication that at least 90\,per\,cent
of high-redshift DLAs may have CNM fractions significantly less than
that measured for the Milky Way has important consequences for the
heating and cooling of neutral gas in the early Universe and star
formation (e.g. \citealt*{Wolfe:2003a}). However, these targeted
observations rely on the limited availability of simultaneous 21\,cm
and optical/UV data for the DLAs and assumes commonality between the
column density probed by the optical and radio sight-lines. The first
issue can be overcome by improving the sample statistics through
larger 21\,cm line surveys of high-redshift DLAs, but the latter
requires improvements to our methodology and understanding of the gas
distribution in these systems. There are also concerns about the
accuracy to which the fraction of the source structure subtended by
the absorber can be measured in each system, which can only be
resolved through spectroscopic very long baseline interferometry
(VLBI). It has been suggested that the observed evolution in spin
temperature could be biased by assumptions about the radio-source
covering factor (\citealt{Curran:2005}) and its behaviour as a
function of redshift (\citealt{Curran:2006b, Curran:2012b}).

In this paper we consider an approach using the statistical constraint
on the average spin temperature achievable with future large 21\,cm
surveys using precursor telescopes to the Square Kilometre Array
(SKA). This will enable independent verification of the evolution in
spin temperature at high redshift and provide a method of studying the
global properties of neutral gas below $z \approx 1.7$, where the
Lyman\,$\alpha$ line is inaccessible using ground-based observatories.
In an early attempt at a genuinely blind 21\,cm absorption survey,
\cite{Darling:2011} used pilot data from the Arecibo Legacy Fast
Arecibo L-band Feed Array (ALFALFA) survey to obtain upper limits on
the column density frequency distribution from 21\,cm absorption at
low redshift ($z \lesssim 0.06$). However, they also noted that the
number of detections could be used to make inferences about the ratio
of the spin temperature to covering factor. Building upon this work,
\cite{Wu:2015} found that their upper limits on the frequency
distribution function measured from the 40\,per\,cent ALFALFA survey
({$\alpha$}.40; \citealt{Haynes:2011}) could only be reconciled with
measurements from other low-redshift 21\,cm surveys if the typical
spin temperature to covering factor ratio was greater than 500\,K. At
higher redshifts, \cite{Gupta:2009} found that the number density of
21\,cm absorbers in known \mbox{Mg\,{\sc ii}} absorbers appeared to
decrease with redshift above $z \sim 1$, consistent with a reduction
in the CNM fraction. We pursue this idea further by investigating
whether future wide-field 21\,cm surveys can be used to measure the
average spin temperature in distant galaxies that are rich in atomic
gas.

\section{The expected number of intervening \mbox{H\,{\sc i}}
  absorbers}\label{section:expected_number}

We estimate the expected number of intervening \mbox{H\,{\sc i}}
systems towards a sample of background radio sources by evaluating the
following integral over all sight-lines
\begin{equation}\label{equation:expected_number}
  \mu =  \iint{f(N_{\rm HI},X)\,\mathrm{d}X\,\mathrm{d}N_{\rm HI}}, 
\end{equation}
where $f(N_{\rm HI}, X)$ is the frequency distribution as a function
of column density ($N_{\rm HI}$) and comoving path length ($X$). We
use the results of recent surveys for 21\,cm emission in nearby
galaxies (e.g. \citealt{Zwaan:2005}) and high-redshift Lyman-$\alpha$
absorption in the Sloan Digitial Sky Survey (SDSS;
e.g. \citealt*{Prochaska:2005}; \citealt{Noterdaeme:2009}), which show
that $f(N_{\rm HI}, X)$ can be parametrized by a gamma function of the
form
\begin{equation}
  f(N_{\rm HI}, X) = \left({f_{\ast} \over N_{\ast}}\right)\left({N_{\rm HI} \over N_{\ast}}\right)^{-\beta}\exp{\left(-{N_{\rm HI} \over N_{\ast}}\right)}\,\mathrm{cm}^{2},
\end{equation}
where $f_{\ast} = 0.0193$, $\log_{10}(N_{\ast}) = 21.2$ and $\beta =
1.24$ at $z = 0$ (\citealt{Zwaan:2005}), and $f_{\ast} = 0.0324$,
$\log_{10}(N_{\ast}) = 21.26$ and $\beta = 1.27$ at $z \approx 3$
(\citealt{Noterdaeme:2009}). While the observational data do not yet
constrain models for evolution of the \mbox{H\,{\sc i}} distribution
at intermediate redshifts between $z \sim 0.1$ and
$3$\footnote{Measurements of $f(N_{\rm HI},X)$ at intermediate
  redshifts come from targeted ultra-violet surveys of DLAs using the
  \emph{Hubble Space Telescope} (\citealt*{Rao:2006};
  \citealt{Neeleman:2016}). However, due to the limited sample sizes
  these are currently an order-of-magnitude less sensitive than the
  nearby 21-cm and high-redshift optical Lyman-$\alpha$ surveys.}, it
is known to be much weaker than the significant decline seen in the
global star-formation rate and molecular gas over the same epoch
(e.g. \citealt{Lagos:2014}). We therefore carry out a simple linear
interpolation between the low and high redshift epochs to estimate
$f(N_{\rm HI},X)$ as a function of redshift.

\begin{figure}
\centering
\includegraphics[width=0.475\textwidth]{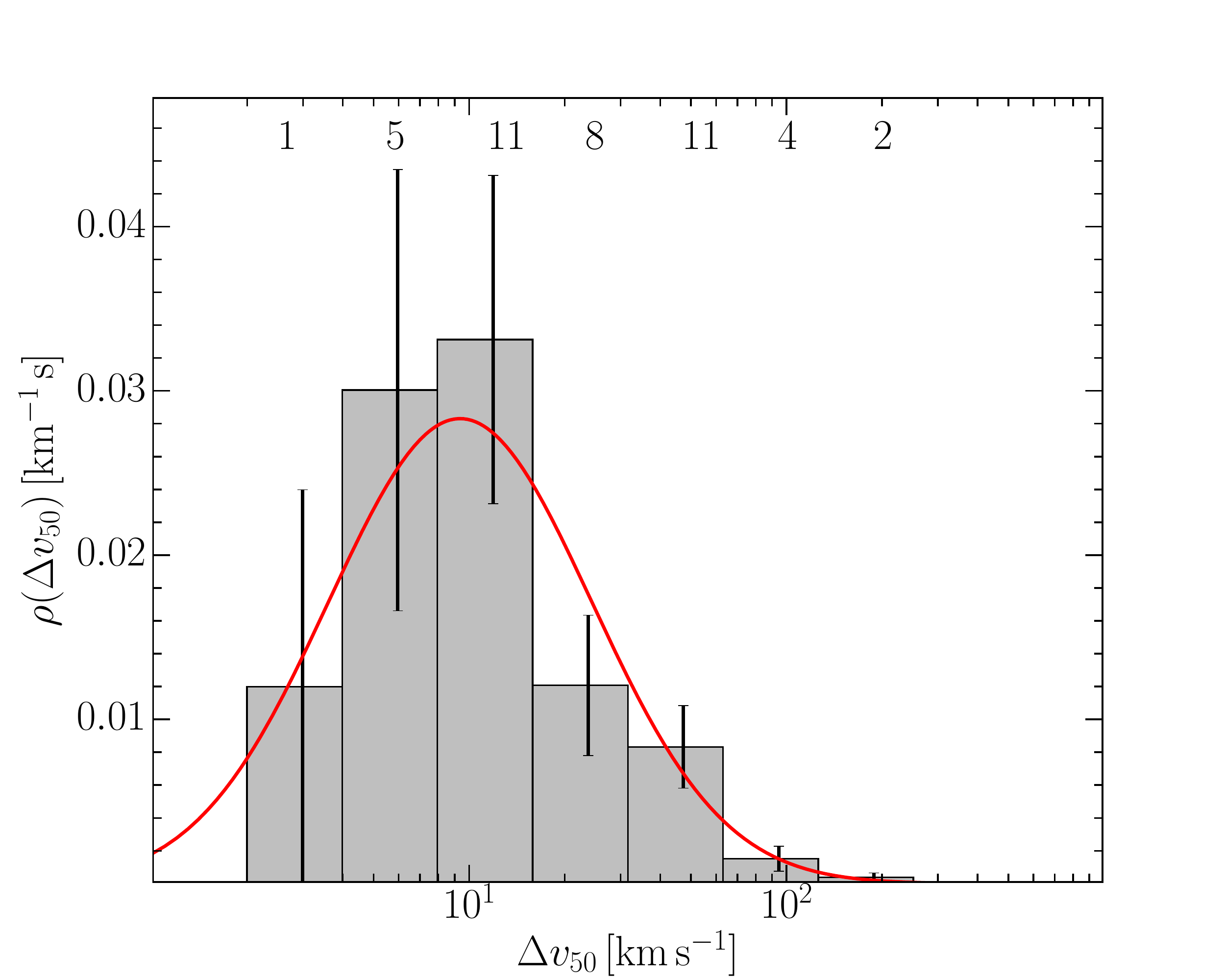}
\caption{The distribution of 21\,cm line widths based on existing
  detections of intervening absorption at $z > 0.1$ (see the text for
  details of this sample). The sample size in each bin is denoted by
  the number above and errorbars denote the standard deviation.  The
  solid red line is a log-normal fit to the data, from which we draw
  random samples for our analysis.}\label{figure:width_dist}
\end{figure}

The probability of detecting an absorbing system of given column
density depends on the sensitivity of the survey, the flux density and
structure of the background source and the fraction of \mbox{H\,{\sc
    i}} in the lower spin state, given by the spin temperature. We
express the column density ($N_{\rm HI}$; in atoms\,cm$^{-2}$) in
terms of the optical depth ($\tau$) and spin temperature ($T_{\rm
  spin}$; in K) by
\begin{equation}\label{equation:column_density}
  N_{\rm HI} = 1.823\times10^{18}\,T_{\rm spin} \int{\tau(v)\mathrm{d}v},
\end{equation}
where the integral is performed across the spectral line in the system
rest-frame velocity $v$ (in km\,s$^{-1}$). We then express the optical
depth in terms of the observables as
\begin{equation}
  \tau = -\ln\left[1 + {\Delta{S}\over c_{\rm f}S_{\rm cont}}\right],
\end{equation}
where $\Delta{S}$ is the observed change in flux density due to
absorption, $S_{\rm cont}$ is the background continuum flux density
and $c_{\rm f}$ is the (often unknown) fraction of background flux
density subtended by the intervening gas.

\begin{figure}
\centering
\includegraphics[width=0.465\textwidth]{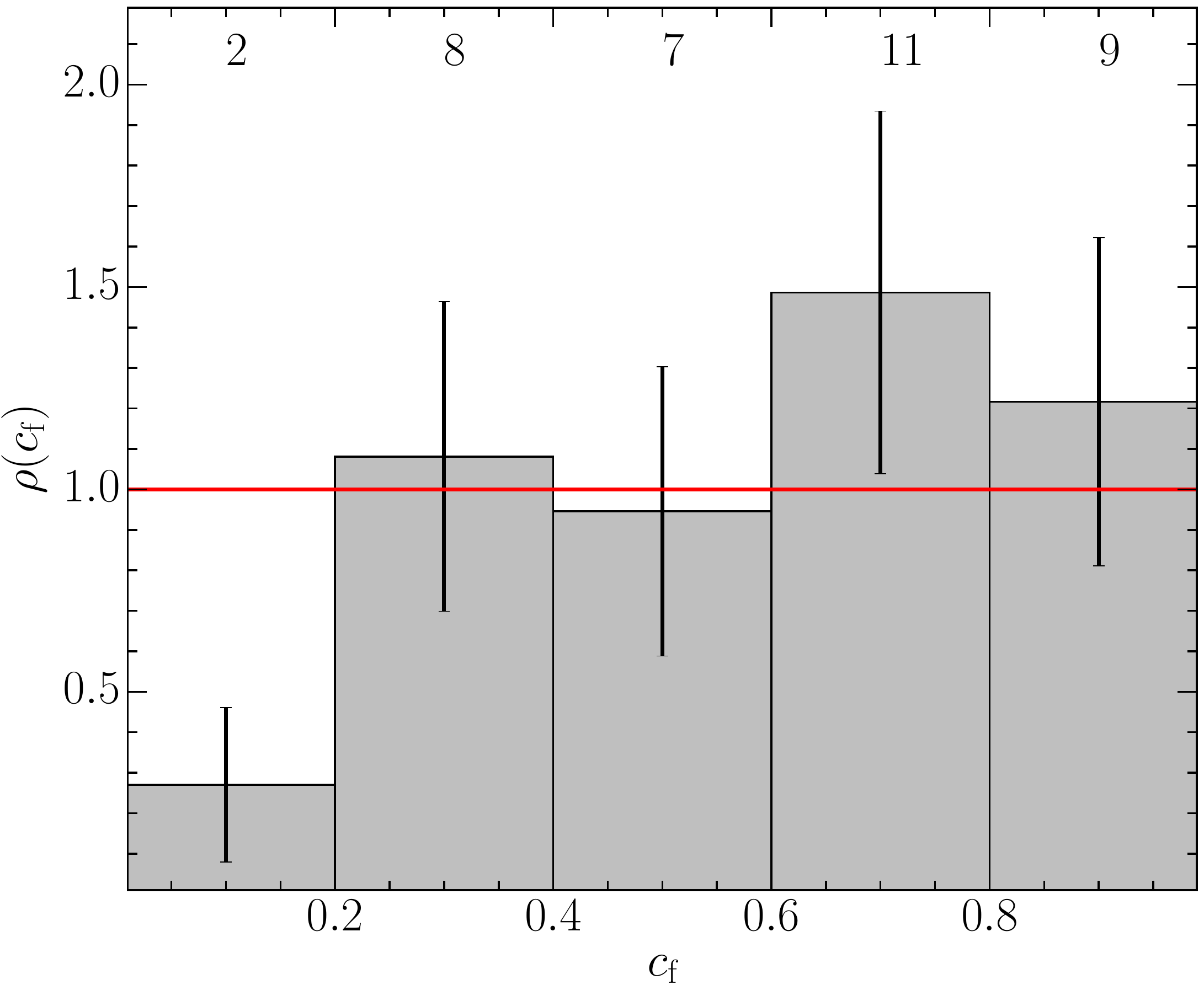}
\caption{The distribution of \mbox{H\,{\sc i}} covering factors from
  the main sample of \citet{Kanekar:2014a}, which were estimated using
  the fraction of total continuum flux density in the quasar core. The
  sample size in each bin is denoted by the number above and errorbars
  denote the standard deviation. The solid red line is the uniform
  distribution, from which we draw random samples for our
  analysis.}\label{figure:covfact_dist}
\end{figure}

We assume that a single intervening system can be described by a
Gaussian velocity distribution of full width at half maximum (FWHM)
dispersion ($\Delta{v_{\rm 50}}$) and peak optical depth ($\tau_{\rm
  peak}$), so that \autoref{equation:column_density} can be re-written
as
\begin{equation}\label{equation:column_density_gaussian}
  N_{\rm HI} = 1.941\times10^{18}\,T_{\rm spin}\,\tau_{\rm peak}\,\Delta{v_{\rm 50}}.
\end{equation}
If we further assume that the rms spectral noise is Gaussian, with a
standard deviation $\sigma_{\rm chan}$ per independent channel
$\Delta{v_{\rm chan}}$, then the 5$\sigma$ column density detection
limit is given by
\begin{equation}
  N_{5\sigma} \approx 1.941\times10^{18}\,T_{\rm spin}\,\tau_{\rm 5\sigma}\,\Delta{v_{\rm conv}},
\end{equation}
where 
\begin{equation}\label{equation:optical_depth_limit}
  \tau_{5\sigma} \approx -\ln\left[1 - {5\,\sigma_{\rm chan}\over c_{\rm f}\,S_{\rm cont}}\sqrt{\Delta{v}_{\rm chan}\over \Delta{v_{\rm conv}}}\right],
\end{equation}
and $\Delta{v_{\rm conv}} \approx \sqrt{\Delta{v}_{\rm chan}^{2} +
  \Delta{v}_{50}^{2}}$, which is the observed width of the line, given
by the convolution of the physical velocity distribution and the
spectral resolution of the telescope. We now redefine $\mu$ as the
expected number of intervening \mbox{H\,{\sc i}} detections in our
survey as a function of the column density sensitivity along each
sight-line where each comoving path element
$\delta{X}(z)$\footnote{For the purposes of this work we adopt a flat
  $\Lambda$ cold dark matter cosmology with $H_{0}$ =
  70\,km\,s$^{-1}$, $\Omega_\mathrm{M}$ = 0.3 and $\Omega_{\Lambda}$ =
  0.7. }  in the integral defined by
\autoref{equation:expected_number} is given by
\begin{equation}
  \delta{X}(z)= 
\begin{cases}
  {\delta{z}\,(1+z)^{2}\over \sqrt{(1+z)^{2}(1+z\Omega_{\rm
        M})-z(z+2)\Omega_{\rm \Lambda}}}, & \text{if}\ N_{\rm HI} \geq N_{5\sigma}, \\
  0, & \text{otherwise}.
\end{cases}
\end{equation}

To calculate the column density sensitivity for each comoving element
we draw random samples for $\Delta{v}_{50}$ and $c_{\rm f}$ from
continuous prior distributions based on existing evidence. In the case
of $\Delta{v}_{50}$ we use a log-normal distribution obtained from a
simple least-squares fit to the sample distribution from previous
21-cm absorption surveys reported in the literature (see
\autoref{figure:width_dist})\footnote{References for the literature
  sample of line widths shown in \autoref{figure:width_dist}:
  \citet*{Briggs:2001}; \citet*{Carilli:1993};
  \citet*{Chengalur:1999}; \citet{Chengalur:2000, Curran:2007b,
    Davis:1978, Ellison:2012, Gupta:2009, Gupta:2012, Gupta:2013};
  \citet{Kanekar:2001b,Kanekar:2003b}; \citet{Kanekar:2001c,
    Kanekar:2006, Kanekar:2009a, Kanekar:2013, Kanekar:2014a};
  \citet{Kanekar:2003a}, \citet*{Kanekar:2007};
  \citet*{Kanekar:2014b}; \citet{ Lane:2001, Lovell:1996, York:2007,
    Zwaan:2015}.}, assuming that this correctly describes the true
distribution for the population of DLAs. However, direct measurement
of the \mbox{H\,{\sc i}} covering factor is significantly more
difficult and so for the purposes of this work we draw random samples
assuming a uniform distribution between 0 and 1. In
\autoref{figure:covfact_dist}, we show a comparison between this
assumption and the sample distribution estimated by
\cite{Kanekar:2014a} from their main sample of 37 quasars. Kanekar et
al. used VLBI synthesis imaging to measure the fraction of total
quasar flux density contained within the core, which was then used as
a proxy for the covering factor. By carrying out a two-tailed
Kolmogorov-Smirnov (KS) test of the hypothesis that the Kanekar et
al. data are consistent with our assumed uniform distribution, we find
that this hypothesis is rejected at the 0.05 level, but not at the
0.01 level (this outcome is dominated by the paucity of quasars in the
sample with $c_{\rm f} \lesssim 0.2$). It is therefore possible that
the population distribution of \mbox{H\,{\sc i}} covering factors may
deviate somewhat from the uniform distribution assumed in this
work. We discuss the implications of this further in
\autoref{section:covering_factor}.

\section{A 21\,cm absorption survey with
  ASKAP}\label{section:all_sky_survey}

We use the Australian Square Kilometre Array Pathfinder (ASKAP;
\citealt{Johnston:2007}) as a case study to demonstrate the expected
results from planned wide-field surveys for 21\,cm absorption
(e.g. the ASKAP First Large Absorption Survey in \mbox{H\,{\sc i}} --
Sadler et al., the MeerKAT Absorption Line Survey -- Gupta et al., and
the Search for HI absorption with AperTIF -- Morganti et al.). ASKAP
is currently undergoing commissioning. Proof-of-concept observations
with the Boolardy Engineering Test Array (\citealt{Hotan:2014}) have
already been used to successfully detect a new \mbox{H\,{\sc i}}
absorber associated with a probable young radio galaxy at $z = 0.44$
(\citealt{Allison:2015a}). Here we predict the outcome of a future
2\,h-per-pointing survey of the entire southern sky ($\delta \leq
+10\degr$) using the full 36-antenna ASKAP in a single 304\,MHz band
between 711.5 and 1015.5\,MHz, equivalent to \mbox{H\,{\sc i}}
redshifts between $z = 0.4$ and 1.0.

\begin{figure}
\centering
\includegraphics[width=0.475\textwidth]{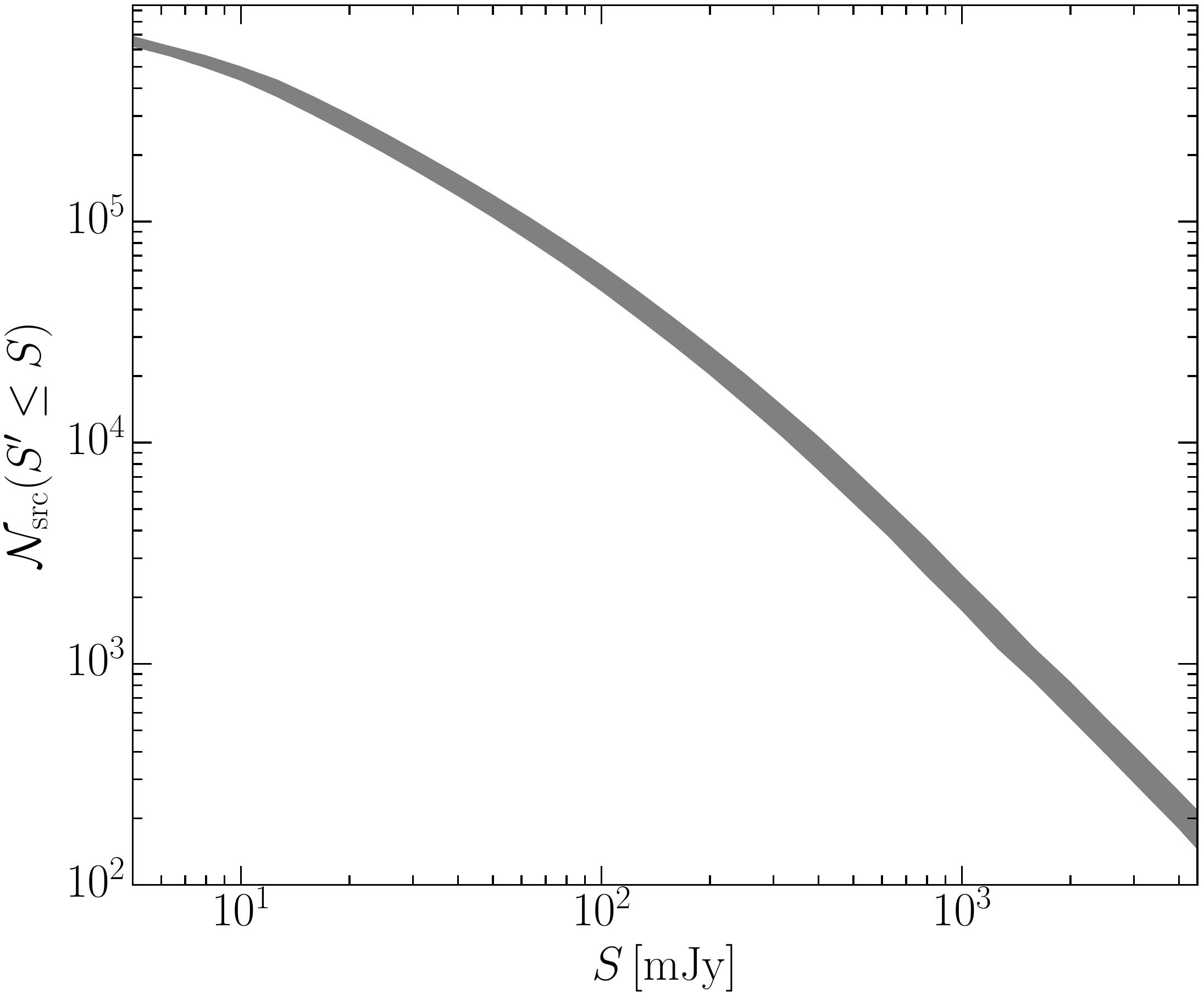}
\caption{The number of radio sources in our simulated southern sky
  survey ($\delta \leq +10\degr$) estimated from existing catalogues
  at $L$-band frequencies (see the text for details). The grey region
  encloses the expected number across the 711.5 - 1015.5\,MHz ASKAP
  frequency band, assuming a canonical spectral index of $\alpha =
  -0.7$.}\label{figure:nsources_flux}
\end{figure}

Our expectations of the ASKAP performance are based on preliminary
measurements by \cite{Chippendale:2015} using the prototype Mark {\sc
  II} phase array feed. We estimate the noise per spectral channel
using the radiometer equation
\begin{equation}
  \sigma_{\rm chan} =  {S_{\rm system} \over \sqrt{n_{\rm pol}\,n_{\rm ant}\,(n_{\rm ant} - 1)\,\Delta{t}_{\rm in}\,\Delta{\nu}_{\rm chan}}}, 
\end{equation}
where $S_{\rm system}$ is the system equivalent flux density, $n_{\rm
  pol}$ is the number of polarizations, $n_{\rm ant}$ is the number of
antennas, $\Delta{t}_{\rm in}$ is the on-source integration time and
$\Delta{\nu}_{\rm chan}$ is the spectral resolution in frequency. The
sensitivity of the telescope in the 711.5 - 1015.5\,MHz band is
expected to vary between $S_{\rm system} \approx 3200$ and $2000$\,Jy,
with the largest change in sensitivity between 700 and 800\,MHz. ASKAP
has dual linear polarization feeds, 36 antennas and a fine filter bank
that produces 16\,416 independent channels across the full 304\,MHz
bandwidth, so the expected noise per 18.5\,kHz channel in a 2\,h
observation is approximately 5.5 - 3.5\,mJy\,beam$^{-1}$ across the
band. In the case of an actual survey, the true sensitivity will of
course be recorded in the spectral data as a function of redshift (see
e.g. \citealt{Allison:2015a}), but for the purposes of the simulated
survey presented in this work we split the band into several frequency
bins to capture the variation in sensitivity and velocity resolution
(which is in the range 7.8\,km\,s$^{-1}$ at 711.5\,MHz to
5.5\,km\,s$^{-1}$ at 1015.5\,MHz).

\begin{figure}
\centering
\includegraphics[width=0.475\textwidth]{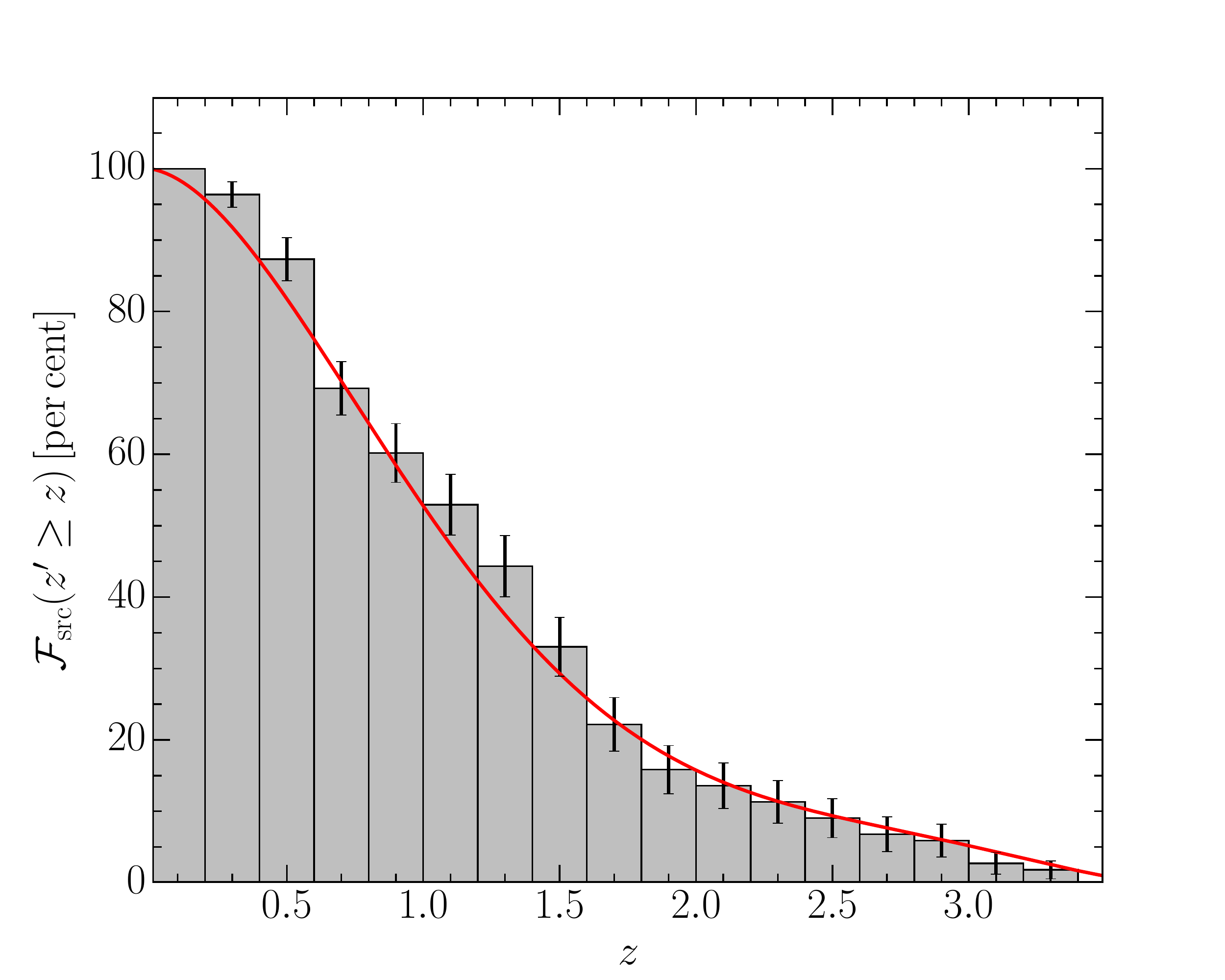}
\caption{The distribution of CENSORS sources (\citealt{Brookes:2008})
  brighter than 10\,mJy beyond a given redshift $z$. The red line
  denotes the cumulative distribution calculated from the parametric
  model of \citet{deZotti:2010}.}\label{figure:zdist}
\end{figure}

In order to simulate a realistic survey of the southern sky we select
all radio sources south of $\delta = +10\degr$ from catalogues of the
National Radio Astronomy Observatory Very Large Array Sky Survey
(NVSS, $\nu = 1.4$\,GHz, $S_{\rm src} \gtrsim 2.5$\,mJy;
\citealt{Condon:1998}), the Sydney University Molonglo SkySurvey ($\nu
= 843$\,MHz, $S_{\rm src} \gtrsim 10$\,mJy; \citealt{Mauch:2003}) and
the second epoch Molonglo Galactic Plane Survey ($\nu = 843$\,MHz,
$S_{\rm src} \gtrsim 10$\,mJy; \citealt{Murphy:2007}). The source flux
densities, used to calculate the optical depth limit in
\autoref{equation:optical_depth_limit}, are estimated at the centre of
each frequency bin by extrapolating from the catalogue values and
assuming a canonical spectral index of $\alpha = -0.7$. In
\autoref{figure:nsources_flux}, we show the resulting cumulative
distribution of radio sources in our sample as a function of flux
density across the band.

\begin{figure}
\centering
\includegraphics[width=0.475\textwidth]{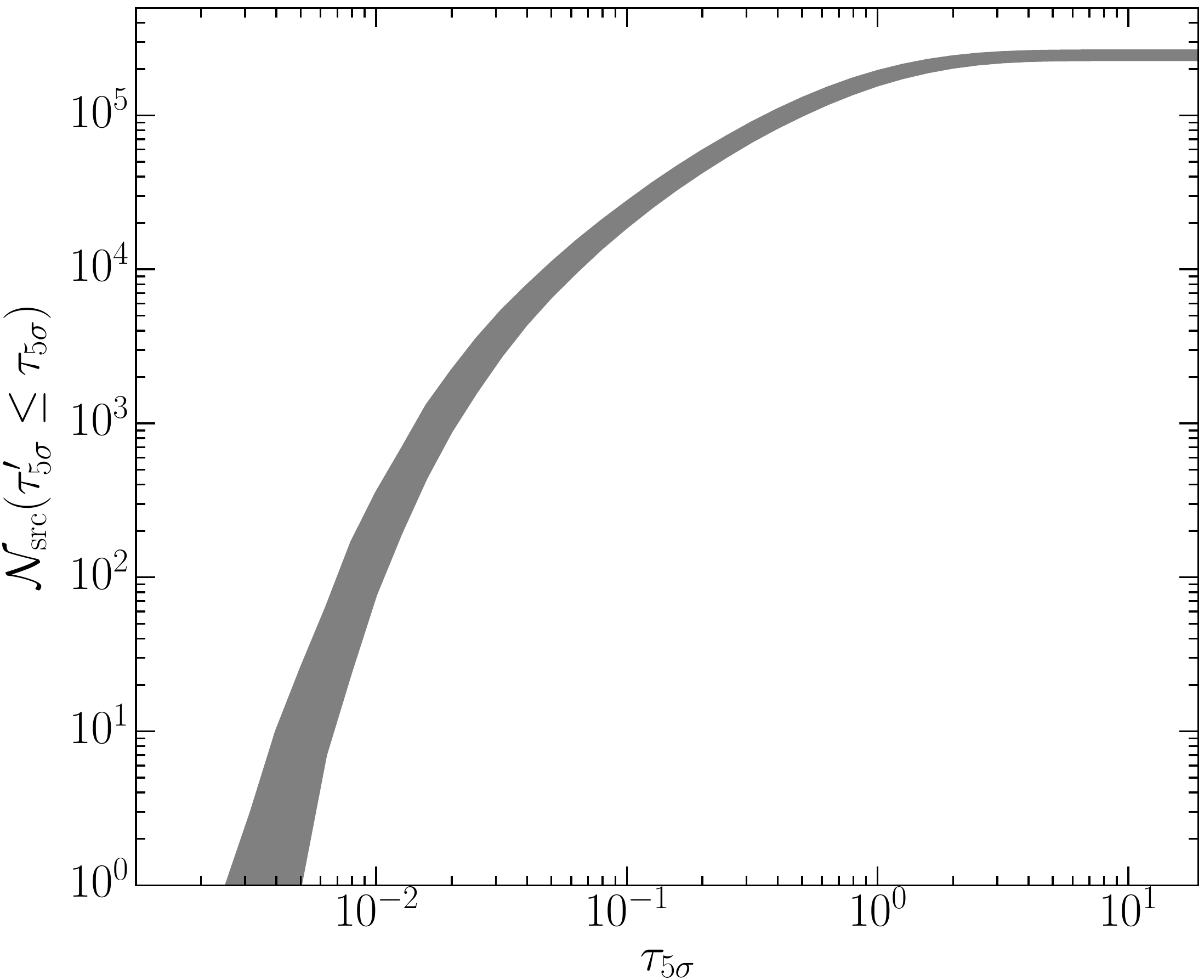}
\caption{The number of sources in our simulated ASKAP survey with a
  21\,cm opacity sensitivity greater than or equal to
  $\tau_{5\sigma}$, as defined by
  \autoref{equation:optical_depth_limit}. The grey region encloses
  opacity sensitivities for the 711.5 - 1011.5\,MHz band. Random
  samples for the line FWHM and covering factor were drawn from the
  distributions shown in \autoref{figure:width_dist} and
  \autoref{figure:covfact_dist}.}\label{figure:nsources_tau}
\end{figure}
  
For any given sight-line, the redshift interval over which absorption
may be detected is dependent upon the distance to the continuum
source. The lack of accurate spectroscopic redshift measurements for
most radio sources over the sky necessitates the use of a statistical
approach based on a model for the source redshift distribution. We
therefore apply a statistical weighting to each comoving path element
$\delta{X}(z)$ such that the expected number of absorber detections is
now given by
\begin{equation}\label{equation:weighted_sum_number}
  \mu =  \iint{f(N_{\rm HI},X)\,\mathcal{F}_{\rm src}(z^{\prime} \geq z)\,\mathrm{d}X\,\mathrm{d}N_{\rm HI}}, 
\end{equation}
where
\begin{equation}
  \mathcal{F}_{\rm src}(z^{\prime} \geq z) = {\int_{z}^{\infty} \mathcal{N}_{\rm src}(z^{\prime})\mathrm{d}z^{\prime}\over\int_{0}^{\infty}\mathcal{N}_{\rm src} (z^{\prime})\mathrm{d}z^{\prime}},
\end{equation}
and $\mathcal{N}_{\rm src}(z)$ is the number of radio sources as a
function of redshift. To estimate $\mathcal{N}_{\rm src}(z)$ we use
the Combined EIS-NVSS Survey Of Radio Sources (CENSORS;
\citealt{Brookes:2008}), which forms a complete sample of radio
sources brighter than 7.2\,mJy at 1.4\,GHz with spectroscopic
redshifts out to cosmological distances. In \autoref{figure:zdist} we
show the distribution of CENSORS sources brighter than 10\,mJy beyond
a given redshift $z$, and the corresponding analytical function
derived from the model fit of \citet{deZotti:2010}, given by
\begin{equation}
  \mathcal{N}_{\rm src}(z) \approx 1.29 + 32.37z - 32.89z^{2} + 11.13z^{3} - 1.25z^{4},
\end{equation}
which we use in our analysis. For the redshifts spanned by our
simulated ASKAP survey, the fraction of background sources evolves
from 87\,per\,cent at $z = 0.4$ to 53\,per\,cent at $z = 1.0$.

We assume that this redshift distribution applies to any sight-line
irrespective of the continuum flux density. However, this assumption
is only true if the source population in the target sample evolves
such that the effect of distance is nullified by an increase in
luminosity. Given this criterion, and the sensitivity of our simulated
survey, we limit our sample to sources with flux densities between 10
and 1000\,mJy, which are dominated by the rapidly evolving population
of high-excitation radio galaxies and quasars
(e.g. \citealt{Jackson:1999, Best:2012, Best:2014, Pracy:2016}) and
for which the redshift distribution is known to be almost independent
of flux density (e.g. \citealt{Condon:1984, Condon:1998}). In
\autoref{figure:nsources_tau}, we show the number of sources from this
sub-sample as a function of opacity sensitivity [as defined by
\autoref{equation:optical_depth_limit}], drawing random samples of the
line FWHM and covering factor from the distributions shown in
\autoref{figure:width_dist} and \autoref{figure:covfact_dist}. There
are approximately 190\,000 sightlines with sufficient sensitivity to
detect absorption of optical depth greater than $\tau_{5\sigma}
\approx 1.0$ and 25\,000 sensitive to optical depths greater than
$\tau_{5\sigma} \approx 0.1$. Since this distribution converges at
optical depth sensitivities greater than $\tau_{5\sigma} \approx 5$,
the population of sources fainter than 10\,mJy, which are excluded
from our simulated ASKAP survey, would not significantly contribute to
further detections of absorption. Similarly, while sources brighter
than 1\,Jy are good probes of low-column-density \mbox{H\,{\sc i}}
gas, they do not constitute a sufficiently large enough population to
significantly affect the total number of absorber detections expected
in the survey and can also be safely excluded.

\begin{figure}
\centering
\includegraphics[width=0.475\textwidth]{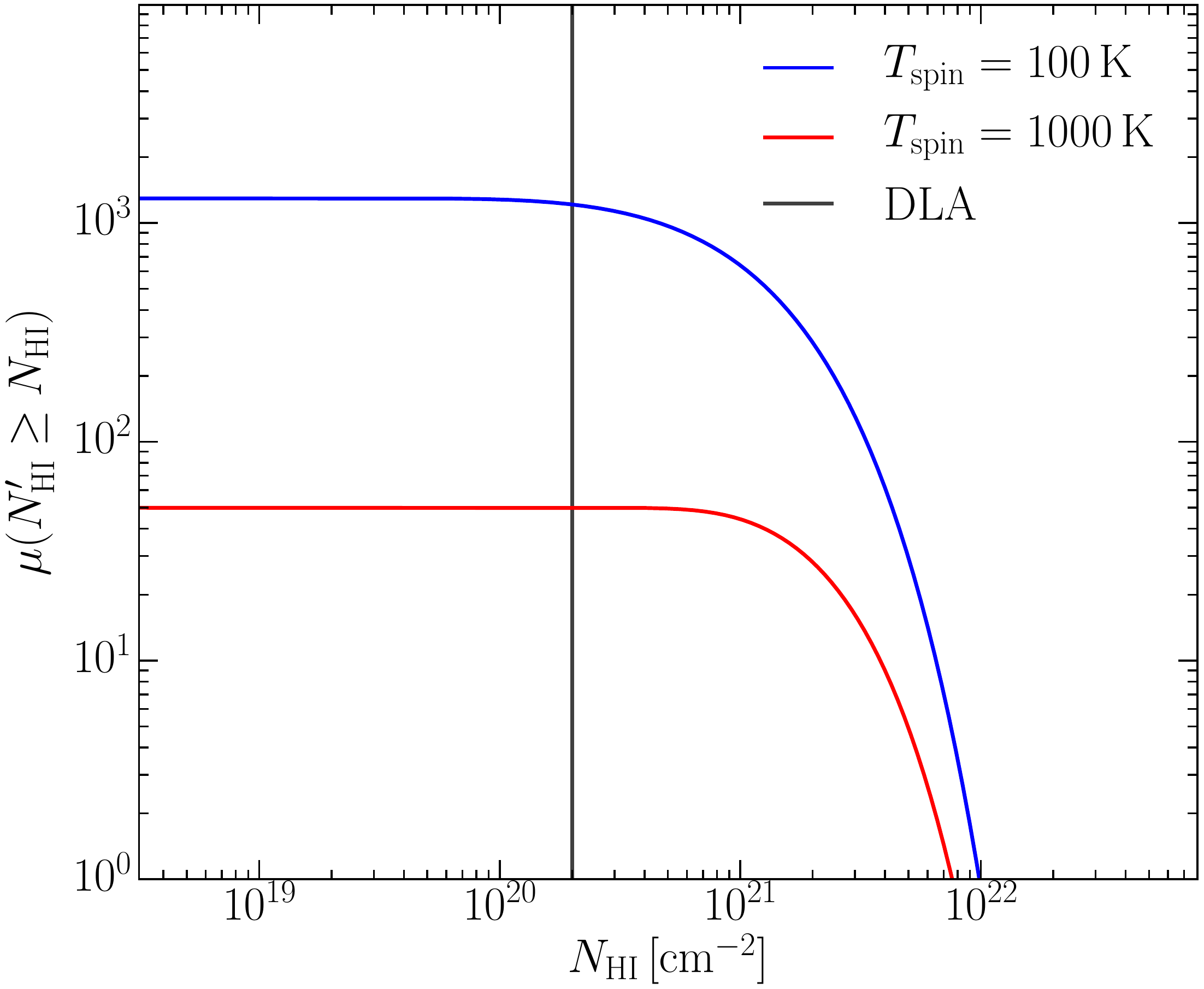}
\caption{The expected number of absorber detections (as a cumulative
  function of column density) in our simulated ASKAP survey. We show
  two scenarios for a single spin temperature $T_{\rm spin} = 100$ and
  $1000$\,K, where we have drawn random samples for the line FWHM and
  covering factor from the distributions shown in
  \autoref{figure:width_dist} and \autoref{figure:covfact_dist}. In
  both cases we find that the expected number of detections is not
  sensitive to column densities below $N_{\rm HI} = 2 \times
  10^{20}$\,cm$^{-2}$, indicating that such a survey will only be
  sensitive to DLA systems.}\label{figure:ndetections_nhi}
\end{figure}

Based on these assumptions, we can estimate the number of absorbers we
would expect to detect in our survey with ASKAP as a function of spin
temperature. In \autoref{figure:ndetections_nhi}, we show the expected
detection yield as a cumulative function of column density. We show
results for two scenarios where the spin temperature is fixed at a
single value of either 100 or 1000\,K, and the line FWHM and covering
factors are drawn from the random distributions shown in
\autoref{figure:width_dist} and \autoref{figure:covfact_dist}. We find
that for both these cases the expected number of detections is not
sensitive to column densities below the DLA definition of $N_{\rm HI}
= 2\times 10^{20}$\,cm$^{-2}$. We also show in
\autoref{figure:ndetections_total} the expected total detection yield
(integrated over all \mbox{H\,{\sc i}} column densities) as a function
of a single spin temperature $T_\mathrm{spin}$ and line FWHM
$\Delta{v}_\mathrm{50}$. We find that for typical spin temperatures of
a few hundred kelvin (consistent with the typical fraction of CNM
observed in the local Universe) and a line FWHM of approximately
$20$\,km\,s$^{-1}$, a wide-field 21\,cm survey with ASKAP is expected
to yield $\sim 1000$ detections. However, even moderate evolution to a
higher spin temperature in the DLA population should see significant
reduction in the detection yield from this survey.

\section{Inferring the average spin
  temperature}\label{section:spin_temp}

\begin{figure}
\centering
\includegraphics[width=0.475\textwidth]{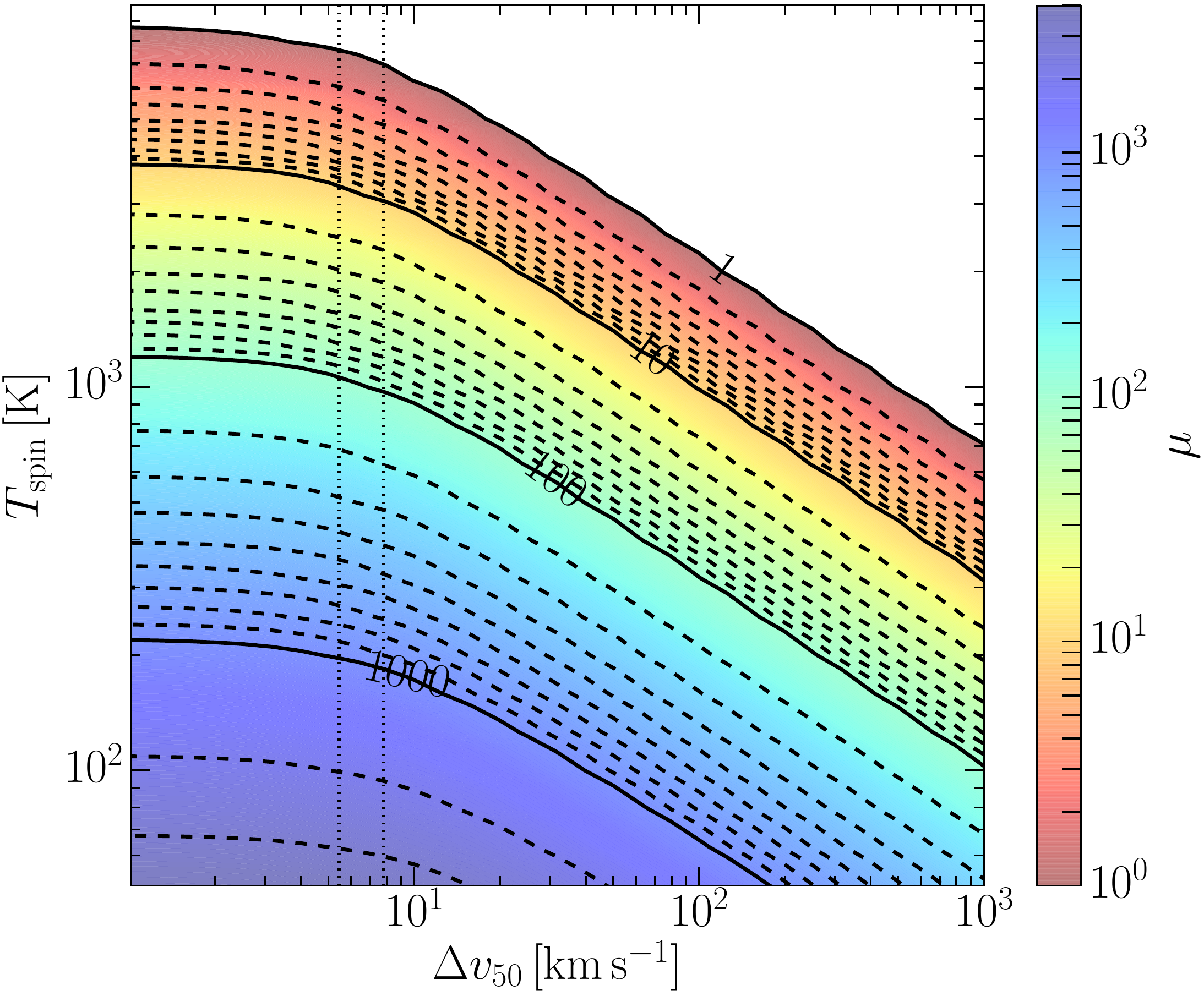}
\caption{The expected total number of detections in our simulated
  ASKAP survey, as a function of a single spin temperature ($T_{\rm
    spin}$) and line FWHM ($\Delta{v}_{50}$). The vertical dotted
  lines enclose the velocity resolution across the observed frequency
  band. We draw random samples for the covering factor from a uniform
  distribution between 0 and 1 (as shown in
  \autoref{figure:covfact_dist}). The contours are truncated at $\mu <
  1$ for clarity.}\label{figure:ndetections_total}
\end{figure}

We cannot directly measure the spin temperatures of individual systems
without additional data from either 21\,cm emission or Lyman-$\alpha$
absorption.  However, from \autoref{figure:ndetections_total} it is
evident that the total number of absorbing systems expected to be
detected with a reasonably large 21\,cm survey is strongly dependent
on the assumed value for the spin temperature. Therefore, by comparing
the actual survey yield with that expected from the known
\mbox{H\,{\sc i}} distribution, we can infer the average spin
temperature of the atomic gas within the DLA population for a given
redshift interval.

Assuming that the total number of detections follows a Poisson
distribution, the probability of detecting $\mathcal{N}$ intervening
absorbing systems is given by
\begin{equation} p(\mathcal{N}|\overline{\mu}) =
  {\overline{\mu}^{\mathcal{N}} \over \mathcal{N}!}
  \mathrm{e}^{-\overline{\mu}},
\end{equation}
where $\overline{\mu}$ is the expected total number of detections
given by the integral
\begin{equation}
  \overline{\mu} = \iiint{\mu(T_{\rm spin},\Delta{v}_{50},c_{\rm f})\rho(T_{\rm spin},\Delta{v}_{50},c_{\rm f})\mathrm{d}T_{\rm spin}\mathrm{d}\Delta{v}_{50}\mathrm{d}c_{\rm f}},
\end{equation}
and $\rho$ is the distribution of systems as a function of spin
temperature, line FWHM and covering factor. We assume that all three
of these variables are independent\footnote{In the case where thermal
  broadening contributes significantly to the velocity dispersion, and
  the spin temperature is dominated by collisional excitation, the
  assumption that these are independent may no longer hold. However,
  given that collisional excitation dominates in the CNM, where
  $T_{\rm spin} \sim 100\,\mathrm{K}$, the velocity dispersion would
  have to satisfy $\Delta{v}_{50} \ll 10$\,km\,s$^{-1}$ (c.f. the
  distribution shown in \autoref{figure:width_dist}).} so that $\rho$
factorizes into functions of each. We then marginalise over the
covering factor and line width distributions shown in
\autoref{figure:width_dist} and \autoref{figure:covfact_dist} so that
the expression for $\overline{\mu}$ reduces to
\begin{equation}\label{equation:harmonic_spin_temp}
  \overline{\mu} = \int{\mu(T_{\rm spin})\rho(T_{\rm spin})\mathrm{d}T_{\rm spin}} = \mu(\overline{T}_{\rm spin}),
\end{equation}
where $\overline{T}_{\rm spin}$ is the harmonic mean of the unknown
spin temperature distribution, weighted by column density. This is
analogous to the spin temperature inferred from the detection of
absorption in a single intervening galaxy averaged over several
gaseous components at different temperatures
(e.g. \citealt{Carilli:1996}).

In the event of the survey yielding $\mathcal{N}$ detections, we can
calculate the posterior probability density of $\overline{T}_{\rm
  spin}$ using the following relationship between conditional
probabilities
\begin{equation}
  p(\overline{T}_{\rm spin}|\mathcal{N}) = {p(\mathcal{N}|\overline{T}_{\rm spin})p(\overline{T}_{\rm
      spin})\over p(\mathcal{N})},
\end{equation}
where $p(\overline{T}_{\rm spin})$ is our prior probability density
for $\overline{T}_{\rm spin}$ and $p(\mathcal{N})$ is the marginal
probability of the number of detections, which can be treated as a
normalizing constant. The minimally informative Jeffreys prior for the
mean value $\mu$ of a Poisson distribution is $1/\sqrt{\mu}$
(\citealt{Jeffreys:1946})\footnote{A suitable alternative choice for
  the prior is the standard scale-invariant form $1/\mu$
  (e.g. \citealt{Jeffreys:1961, Novick:1965, Villegas:1977}). While we
  find that our choice of non-informative prior has negligible effect
  on the spin temperature posterior for the full \mbox{H\,{\sc i}}
  absorption survey, as one would expect this choice becomes more
  important for smaller surveys. For the early-science 1000\,deg$^{2}$
  survey discussed in \autoref{section:tspin_results} we find that the
  difference in these two priors produces a $\sim 2$ to 20\,per\,cent
  effect in the posterior. However, in all cases considered this
  change is smaller than the 68.3\,per\,cent credible interval spanned
  by the posterior.}. From \autoref{equation:harmonic_spin_temp} it
therefore follows that a suitable form for the non-informative spin
temperature prior is $p(\overline{T}_{\rm spin}) =
1/\sqrt{\overline{\mu}}$, so that
\begin{equation}\label{equation:tspin_prob}
  p(\overline{T}_{\rm spin}|\mathcal{N}) = C^{-1}\,{\overline{\mu}^{(\mathcal{N}-1/2)} \over \mathcal{N}!} \mathrm{e}^{-\overline{\mu}}, 
\end{equation}
where the distribution is normalised to unit total probability by
evaluating the integral
\begin{equation}
  C = \int{{\overline{\mu}^{(\mathcal{N}-1/2)} \over \mathcal{N}!} \mathrm{e}^{-\overline{\mu}}}\,\mathrm{d}\overline{T}_{\rm spin}.
\end{equation}

The probabilistic relationship given by \autoref{equation:tspin_prob}
and the expected detection yield derived in
\autoref{section:all_sky_survey} can be used as a frame-work for
inferring the harmonic-mean spin temperature using the results of any
homogeneous 21-cm survey. We have assumed that we can accurately
distinguish intervening absorbing systems from those associated with
the host galaxy of the radio source. However, any 21-cm survey will be
accompanied by follow-up observations, at optical and sub-mm
wavelengths, which will aid identification. Furthermore, future
implementation of probabilistic techniques to either use photometric
redshift information or distinguish between line profiles should
provide further disambiguation.

Of course we have not yet accounted for any error in our estimate of
$\overline{\mu}$, which will increase our uncertainty in
$\overline{T}_\mathrm{spin}$. In the following section we discuss
these possible sources of error and their effect on the result.

\section{Sources of error}\label{section:errors}

Our estimate of the expected number of 21\,cm absorbers is dependent
upon several distributions describing the properties of the foreground
absorbing gas and the background source distribution. For future
large-scale 21\,cm surveys, the accuracy to which we can infer the
harmonic mean of the spin temperature distribution will eventually be
limited by the accuracy to which we can measure these other
distributions. In this section, we describe these errors and their
propagation through to the estimate of $\overline{T}_\mathrm{spin}$,
summarizing our results in \autoref{table:tspin_uncertainties}.

\subsection{The covering factor}\label{section:covering_factor}

\subsubsection{Deviation from a uniform distribution between 0 and 1}

The fraction $c_{\rm f}$ by which the foreground gas subtends the
background radiation source is difficult to measure directly and is
thereby a significant source of error for 21\,cm absorption
surveys. In this work, we have assumed a uniform distribution for
$c_{\rm f}$, taking random values between 0 and 1. In
\autoref{section:expected_number}, we tested this assumption by
comparing it with the distribution of flux density core fractions in a
sample of 37 quasars, used by \cite{Kanekar:2014a} as a proxy for the
covering factor. By carrying out a two-tailed KS test, we found some
evidence (at the 0.05 level) that this quasar sample was inconsistent
with our assumption of a uniform distribution between 0 and
1. Noticeably there seems to be an under-representation of quasars in
the Kanekar et al. sample with estimated $c_{\rm f} \lesssim 0.2$. In
the low optical depth limit, the detection rate is dependent on the
ratio of spin temperature to covering factor, in which case a
fractional deviation in $c_{\rm f}$ will propagate as an equal
fractional deviation in $\overline{T}_\mathrm{spin}$. Based on the
difference seen in the covering factor distribution of the Kanekar et
al. sample and the uniform distribution, we assume that the spin
temperature can deviate by as much as $\pm$10\,per\,cent.

\subsubsection{Evolution with redshift}

We also consider that the covering factor distribution may evolve with
redshift, which would mimic a perceived evolution in the average spin
temperature. Such an effect was proposed by \cite{Curran:2006b} and
\cite{Curran:2012b}, who claimed that the relative change in
angular-scale behaviour of absorbers and radio sources between low-
and high-redshift samples could explain the apparent evolution of the
spin temperature found by \cite{Kanekar:2003b}. To test for this
effect in their larger DLA sample, \cite{Kanekar:2014a} considered a
sub-sample at redshifts greater than $z = 1$, for which the relative
evolution of the absorber and source angular sizes should be
minimal. While the significance of their result was reduced by
removing the lower redshift absorbers from their sample, they still
found a difference at $3.5\,\sigma$ significance between spin
temperature distributions in the two DLA sub-samples separated by a
median redshift of $z = 2.683$.

\begin{figure}
\centering
\includegraphics[width=0.475\textwidth]{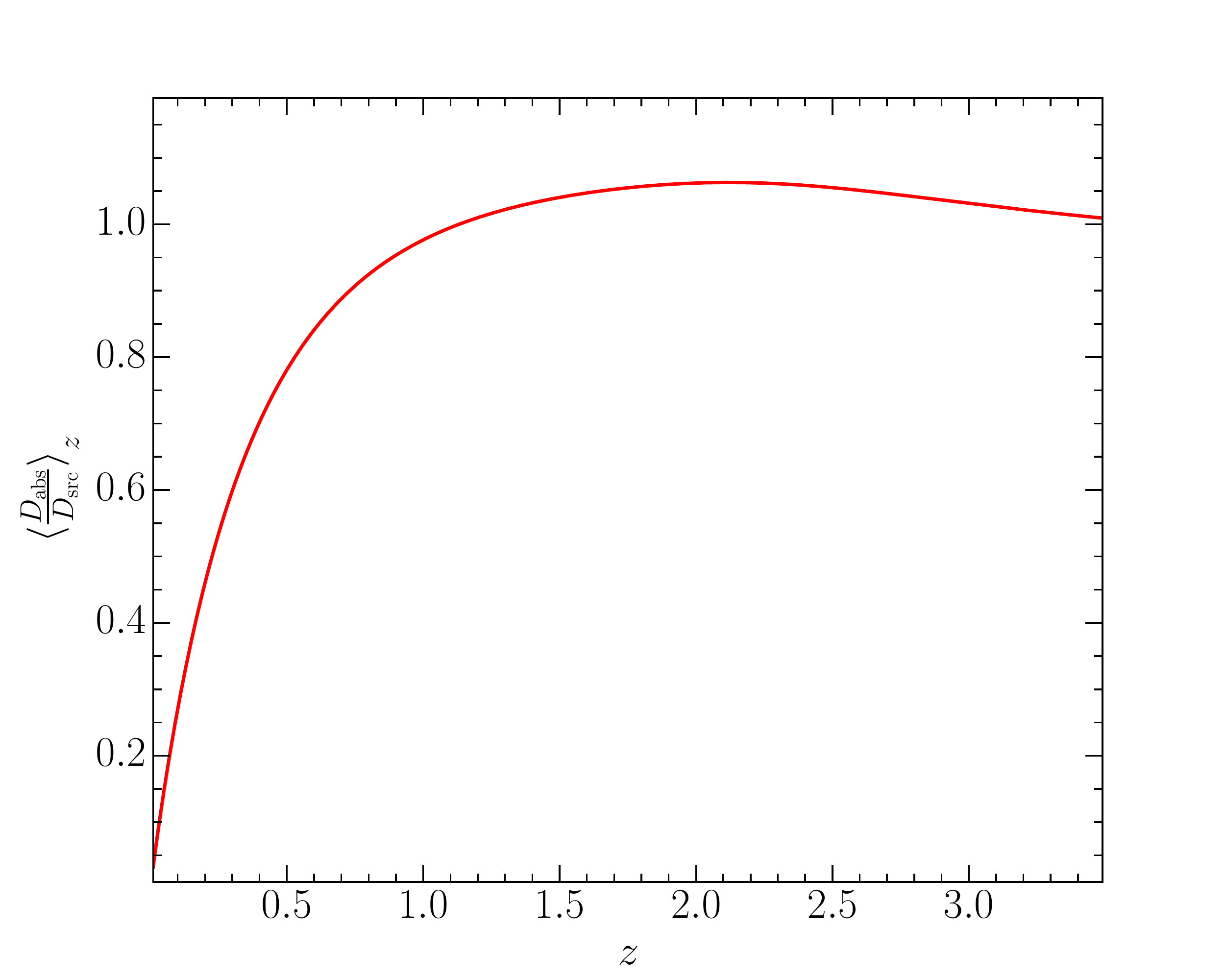}
\caption{The expected redshift behaviour of $D_{\rm abs}/D_{\rm src}$
  based on the \citet{deZotti:2010} model for the radio source redshift
  distribution.}\label{figure:dang_ratio}
\end{figure}

Future surveys with ASKAP and the other SKA pathfinders will search
for \mbox{H\,{\sc i}} absorption at intermediate redshifts ($z \sim
1$), where the relative evolution of the absorber and source angular
sizes is expected to be more significant than for the higher redshift
DLA sample considered by \cite{Kanekar:2014a}. We therefore consider
the potential effect of this cosmological evolution on the inferred
value of $\overline{T}_\mathrm{spin}$. We approximate the covering
factor using the following model of \cite{Curran:2006b}
\begin{equation}\label{equation:covering_factor}
  c_{f} \approx
\begin{cases}
  \left({\theta_{\rm abs}\over \theta_{\rm src}}\right)^{2}, & \text{if}\ \theta_{\rm abs} < \theta_{\rm src}, \\
  1, & \text{otherwise},
\end{cases}
\end{equation}
where $\theta_{\rm abs}$ and $\theta_{\rm src}$ are the angular sizes
of the absorber and background source, respectively. Under the
small-angle approximation $\theta_{\rm abs} \approx {d_{\rm
    abs}/D_{\rm abs}}$ and $\theta_{\rm abs} \approx {d_{\rm
    src}/D_{\rm src}}$, where $d_{\rm abs}$ and $D_{\rm abs}$ are the
linear size and angular diameter distance of the absorber, and
likewise $d_{\rm src}$ and $D_{\rm src}$ are the linear size and
angular diameter distance of the background source. Assuming that the
ratio $d_{\rm abs}/d_{\rm src}$ is randomly distributed and
independent of redshift, any evolution in the covering factor is
therefore dominated by relative changes in the angular diameter
distances. We calculate the expected angular diameter distance ratio
at a redshift $z$ by
\begin{equation}
  \left\langle{D_{\rm abs}\over D_{\rm src}}\right\rangle_{z} = D_{\rm abs}(z){\int_{z}^{\infty} \mathcal{N}_{\rm src}(z^{\prime})D_{\rm src}(z^{\prime})^{-1}\mathrm{d}z^{\prime}\over\int_{z}^{\infty}\mathcal{N}_{\rm src} (z^{\prime})\mathrm{d}z^{\prime}},
\end{equation}
which, for the source redshift distribution model given by
\cite{deZotti:2010}, evolves from 0.7 at $z = 0.4$ to 1.0 at $z = 1.0$
(see \autoref{figure:dang_ratio}).  We note that this is consistent
with the behaviour measured by \cite{Curran:2012b} for the total
sample of DLAs observed at 21\,cm wavelengths. By applying this as a
correction to the otherwise uniformly distributed covering factor
(using \autoref{equation:covering_factor}), we find that the inferred
value of $\overline{T}_{\rm spin}$ systematically increases by
approximately 30 per\,cent.

\subsection{The $\bmath{N_{\rm HI}}$ frequency distribution}

\subsubsection{Uncertainty in the measurement of $f(N_{\rm HI},X)$}

We assume that $f(N_{\rm HI}, X)$ is relatively well understood as a
function of redshift by interpolating between model gamma functions
fitted to the distributions at $z = 0$ and $3$. However, these
distributions were measured from finite samples of galaxies, which of
course have associated uncertainties that need to be considered. In
the case of the data presented by \cite{Zwaan:2005} and
\cite{Noterdaeme:2009}, both have typical measurement uncertainties in
$f(N_{\rm HI}, X)$ of approximately 10\,per\,cent over the range of
column densities for which our simulated ASKAP survey is sensitive
(see \autoref{figure:ndetections_nhi}). This will propagate as a
10\,per\,cent fractional error in the expected number of absorber
detections, and contribute a similar percentage uncertainty in the
inferred average spin temperature.

\subsubsection{Correcting for 21\,cm self-absorption}

In the local Universe, \cite{Braun:2012} showed that self-absorption
from opaque \mbox{H\,{\sc i}} clouds identified in high-resolution
images of the Local Group galaxies M31, M33 and the Large Magellanic
Cloud may necessitate a correction to the local atomic mass density of
up to 30\,per\,cent. Although it is not yet clear whether this small
sample of Local Group galaxies is representative of the low-redshift
population, it is useful to understand how this effect might propagate
through to our average spin temperature measurement. We therefore
replace the gamma-function parametrization of the local $f(N_{\rm
  HI})$ given by \cite{Zwaan:2005} with the non-parametric values
given in table\,2 of \cite{Braun:2012}, and recalculate
$\overline{T}_{\rm spin}$. For an all-sky survey with the full
36-antenna ASKAP we find that $\overline{T}_{\rm spin}$ increases by
$\sim$30 for 100 detections and $\sim$10\,per\,cent for 1000
detections. Note that the correction increases for low numbers of
detections, which are dominated by the highest column density systems.

\subsubsection{Dust obscuration bias in optically-selected DLAs}

At higher redshifts, it is possible that the number density of
optically-selected DLAs could be significantly underestimated as a
result of dust obscuration of the background quasar
(\citealt{Ostriker:1984}). This would cause a reduction in the
$f(\mbox{H\,{\sc i}}, X)$ measured from optical surveys, thereby
significantly underestimating the expected number of intervening
21\,cm absorbers at high redshifts. The issue is further compounded by
the expectation that the highest column density DLAs ($N_{\rm HI}
\gtrsim 10^{21}$\,cm$^{-2}$), for which future wide-field 21\,cm
surveys are most sensitive (see \autoref{figure:ndetections_nhi}), may
contain more dust than their less-dense counterparts.

This conclusion was supported by early analyses of the existing quasar
surveys at that time (e.g. \citealt{Fall:1993}), which indicated that
up to 70\,per\,cent of quasars could be missing from optical surveys
through the effect of dust obscuration, albeit with large
uncertainties. However, subsequent optical and infrared observations
of radio-selected quasars (e.g. \citealt{Ellison:2001};
\citealt*{Ellison:2005}; \citealt{Jorgenson:2006}), which are free of
the potential selection biases associated with these optical surveys,
found that the severity of this issue was substantially over-estimated
and that there was minimal evidence in support of a correlation
between the presence of DLAs and dust reddening. Furthermore, the
\mbox{H\,{\sc i}} column density frequency distribution measured by
\cite{Jorgenson:2006} was found to be consistent with the
optically-determined gamma-function parametrization of
\cite{Prochaska:2005}, with no evidence of DLA systems missing from
the SDSS sample at a sensitivity of $N_{\rm HI} \lesssim 5 \times
10^{21}$\,cm$^{-2}$. Although radio-selected surveys of quasars are
free of the selection biases associated with optical surveys, they do
typically suffer from smaller sample sizes and are therefore less
sensitive to the rarer DLAs with the highest column densities.

Another approach is to directly test whether optically-selected
quasars with intervening DLAs, selected from the SDSS sample, are
systematically more dust reddened than a control sample of non-DLA
quasars. Comparisons in the literature are based on several different
colour indicators, which include the spectral index
(e.g. \citealt{Murphy:2004,Murphy:2016}), spectral stacking
(e.g. \citealt{Frank:2010, Khare:2012}) and direct photometry
(e.g. \citealt*{Vladilo:2008}; \citealt{Fukugita:2015}). The current
status of these efforts is summarized by \citet{Murphy:2016}, showing
broad support for a missing DLA population at the level of
$\sim$5\,per\,cent but highlighting that tension still exists between
different dust measurements. No substantial evidence has yet been
found to support a correlation between the dust reddening and
\mbox{H\,{\sc i}} column density in these optically selected DLA
surveys (e.g. \citealt{Vladilo:2008, Khare:2012, Murphy:2016}).

In an attempt to reconcile the differences and myriad biases
associated with these techniques, \cite{Pontzen:2009} carried out a
statistically-robust meta-analysis of the available optical and radio
data, using a Bayesian parameter estimation approach to model the dust
as a function of column density and metallicity. They found that the
expected fraction of DLAs missing from optical surveys is
7\,per\,cent, with fewer than 28\,per\,cent missing at 3\,$\sigma$
confidence. Based on this body of work we therefore assume that
approximately 10\,per\,cent of DLAs are missing from the SDSS sample
of \cite{Noterdaeme:2009} and consider the affect on our estimate of
$\overline{T}_{\rm spin}$. We further assume that there is no
dependance on column density, an assumption which is supported by the
aforementioned observational data for the range of column densities to
which our 21\,cm survey is sensitive. We find that increasing the
high-redshift column density frequency distribution by 10\,per\,cent
introduces a systematic increase of approximately 3\,per\,cent in the
expected number of detections for the redshifts covered by our ASKAP
surveys. We note that this error will increase significantly for
21\,cm surveys at higher redshifts where the optically derived
$f(\mbox{H\,{\sc i}}, X)$ dominates the calculation of the expected
detection rate.

\subsection{The radio source background}

As described in \autoref{section:all_sky_survey}, we weight the
comoving path-length for each sight-line by a statistical redshift
distribution in order to account for evolution in the radio source
background. We use the parametric model of \cite{deZotti:2010}, which
is derived from fitting the measured redshifts of \cite{Brookes:2008}
for CENSORS sources brighter than 10\,mJy, and assume that this
applies to all sources in the range 10 - 1000\,mJy. In
\autoref{figure:zdist}, we show the cumulative distribution of sources
located behind a given redshift and the associated measurement
uncertainty given by the errorbars. For the intermediate redshifts
covered by the ASKAP survey, the fractional uncertainty in this
distribution increases from $\sigma_{\mathcal{F}_{\rm
    src}}/\mathcal{F}_{\rm src} \approx 3.5$ to 8\,per\,cent between
$z = 0.4$ and 1.0, which propagates through to a similar fractional
uncertainty in $\overline{T}_{\rm spin}$. However, for higher
redshifts this fractional uncertainty increases rapidly at $z > 2$, to
more than 50\,per\,cent at $z = 3$, reflecting the paucity of optical
spectroscopic data for the high-redshift radio source
population. Understanding how the radio source population is
distributed at lower flux densities and at higher redshifts is
therefore a concern for the future 21\,cm absorption surveys
undertaken with the SKA mid- and low-frequency telescopes (see
\citealt{Kanekar:2004} and \citealt*{Morganti:2015} for reviews).

\begin{table} 
 \begin{threeparttable}
   \caption{An account of errors in our estimate of $\overline{T}_{\rm
       spin}$ due to the accuracy to which we can determine the
     expected number of absorber
     detections.}\label{table:tspin_uncertainties}
  \begin{tabular}{l@{\hspace{0.05in}}l@{\hspace{0.05in}}l@{\hspace{0.05in}}l@{\hspace{0.05in}}l} 
    \hline
    &  Source of error &  $\mathrm{err}(\overline{T}_{\rm spin})$ & Refs. \\
    & & [per\,cent] & \\
    \hline
    Covering factor & Distribution uncertainty & $\pm10$ & $a$ \\
    Covering factor & Systematic evolution & +30 & $a$, $b$\\ 
    $f(N_{\rm HI}, X)$ & Measurement uncertainty & $\pm10$ & $c, d$\\
    Low-$z$ $f(N_{\rm HI}, X)$ & Systematic self-absorption & $+(10-30)$ & $e$ \\ 
    High-$z$ $f(N_{\rm HI}, X)$ & Systematic dust-obscuration & $+3$ & $f$, $g$ \\ 
    $\mathcal{F}_{\rm src}(z^{\prime} \geq z)$ & Measurement uncertainty & $\pm 5$ & $h$, $i$ \\
    \hline
   \end{tabular}
   \begin{tablenotes}
    \item[] References: $^{a}${\citet{Kanekar:2014a}},
     $^{b}${\citet{Curran:2012b}}, $^{c}${\citet{Zwaan:2005}},
     $^{d}${\citet{Noterdaeme:2009}} , $^{e}${\citet{Braun:2012}},
     $^{f}${\citet{Pontzen:2009}}, $^{g}${\citet{Murphy:2016}},
     $^{h}${\citet{Brookes:2008}}, $^{i}${\citet{deZotti:2010}}.
   \end{tablenotes}
\end{threeparttable}
\end{table}

\section{Expected results for future 21-cm absorption
  surveys}\label{section:tspin_results}

\begin{figure}
\centering
\includegraphics[width=0.475\textwidth]{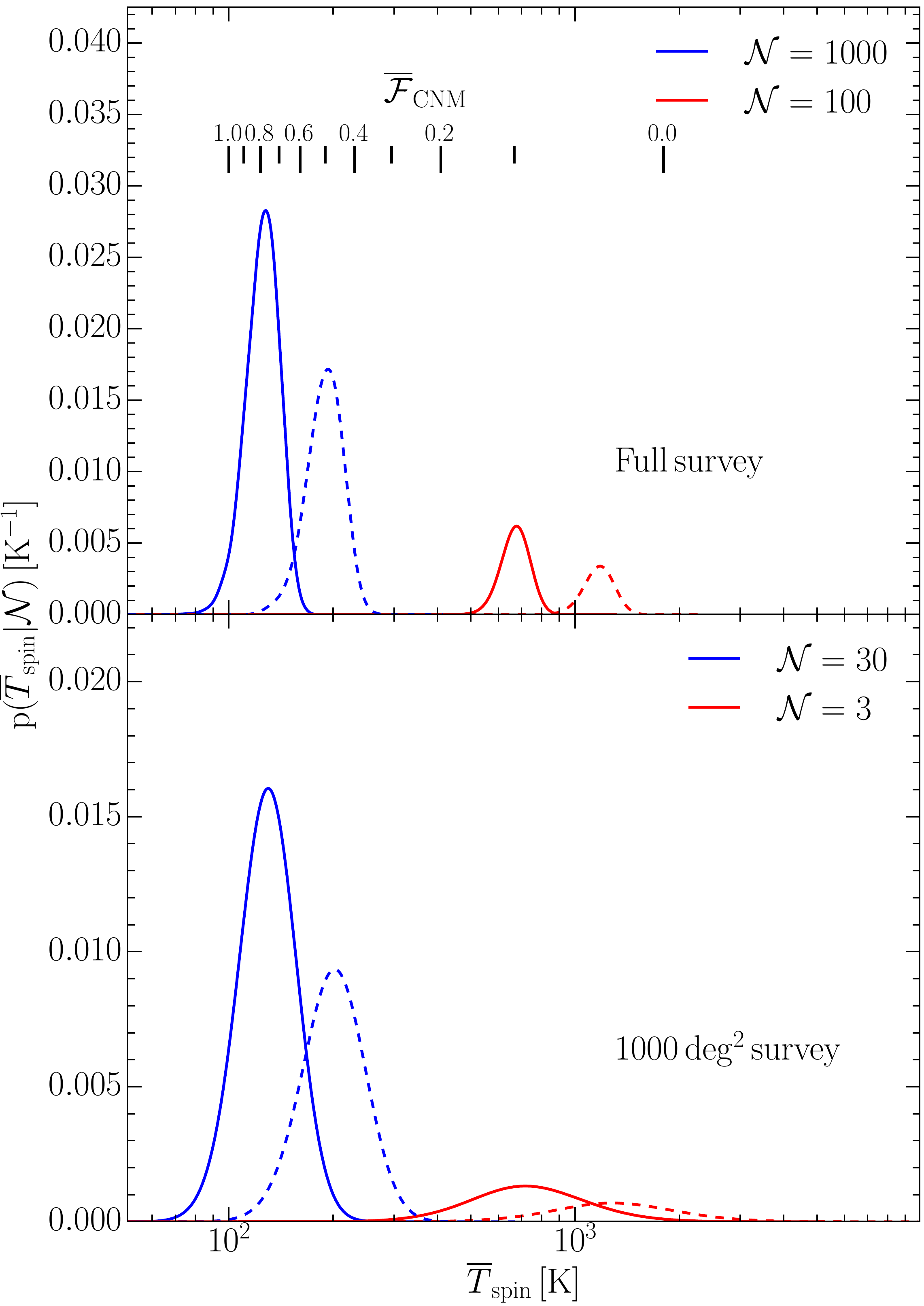}
\caption{The posterior probability density of the average spin
  temperature, as a function of absorber detection yield
  ($\mathcal{N}$). We show results for our simulated all-southern-sky
  survey with 2-h per pointing using the full 36-antenna ASKAP (top
  panel) and a smaller 1000\,deg$^{2}$ survey with 12-h per pointing
  and 12 antennas of ASKAP (bottom panel). The dashed curves show the
  cumulative effect of the systematic errors discussed in
  \autoref{section:errors}. $\overline{\mathcal{F}}_{\rm CNM}$ is the
  average CNM fraction assuming a simple two-phase neutral ISM with
  $T_{\rm spin,CNM} = 100$\,K and $T_{\rm spin,WNM} = 1800$\,K
  (\citealt{Liszt:2001}).}\label{figure:tspin_prob}
\end{figure}

In the top panel of \autoref{figure:tspin_prob} we show the results of
applying our method for inferring $\overline{T}_{\rm spin}$ to the
simulated all-southern-sky \mbox{H\,{\sc i}} absorption survey with
ASKAP described in \autoref{section:all_sky_survey}. We account for
the uncertainties in the expected detection rate $\overline{\mu}$,
discussed in \autoref{section:errors}, by using a Monte Carlo approach
and marginalizing over many realizations. A yield of 1000 absorbers
from such a survey would imply an average spin temperature of
$\overline{T}_\mathrm{spin} =
127^{+14}_{-14}\,(193^{+23}_{-23})$\,K\footnote{We give the
  68.3\,per\,cent interval about the median value measured from the
  posterior distributions shown in \autoref{figure:tspin_prob}.},
where values in parentheses denote the alternative posterior
probability resulting from the systematic errors discussed in
\autoref{section:errors}. This scenario would indicate that a large
fraction of the atomic gas in DLAs at these intermediate redshifts is
in the classical stable CNM phase. Conversely, a yield of only 100
detections would imply that $\overline{T}_\mathrm{spin} =
679^{+64}_{-65}\,(1184^{+116}_{-120})$\,K, indicating that less than
10\,per\,cent of the atomic gas is in the CNM and that the bulk of the
neutral gas in galaxies is significantly different at intermediate
redshifts compared with the local Universe.

We also consider the effect of reducing the sky area and array size,
which is relevant for planned early science surveys with ASKAP and
other SKA pathfinder telescopes. In the bottom panel of
\autoref{figure:tspin_prob}, we show the spin temperatures inferred
when observing a random 1000\,deg$^{2}$ field for 12\,h per pointing,
between $z_{\rm HI} = 0.4$ and $1.0$, using a 12-antenna version of
ASKAP. We find that detection yields of 30 and 3 from such a survey
would give inferred spin temperatures of $\overline{T}_{\rm spin}
=134^{+23}_{-27}\,(209^{+40}_{-47})$ and
$848^{+270}_{-430}\,(1535^{+513}_{-837})$\,K, respectively. The
significant reduction in telescope sensitivity and sky-area,
compensated by the increase in integration time per pointing planned
for early-science, results in a factor of 30 decrease in the expected
number of detections and therefore an increase in the sample variance
and uncertainty in $\overline{T}_{\rm spin}$. However, this result
demonstrates that we expect to be able to distinguish between the
limiting cases of CNM-rich or deficient DLA populations even during
the early-science phases of the SKA pathfinders. For example 30
detections with the early ASKAP survey rules out an average spin
temperature of 1000\,K at high probability.

\section{Conclusions}

We have demonstrated a statistical method for measuring the average
spin temperature of the neutral ISM in distant galaxies, using the
expected detection yields from future wide-field 21\,cm absorption
surveys. The spin temperature is a crucial property of the ISM that
can be used to determine the fraction of the cold ($T_{\rm k} \sim
100$\,K) and dense ($n \sim 100$\,cm$^{-2}$) atomic gas that provides
sites for the future formation of cold molecular gas clouds and star
formation. Recent 21\,cm surveys for \mbox{H\,{\sc i}} absorption in
\mbox{Mg\,{\sc ii}} absorbers and DLAs towards distant quasars have
yielded some evidence of an evolution in the average spin temperature
that might reveal a decrease in the fraction of cold dense atomic gas
at high redshift (e.g. \citealt{Gupta:2009, Kanekar:2014a}).

By combining recent specifications for ASKAP, with available
information for the population of background radio sources, we show
that strong statistical constraints (approximately $\pm10$\,per\,cent)
in the average spin temperature can be achieved by carrying out a
shallow 2-h per pointing survey of the southern sky between redshifts
of $z = 0.4$ and $1.0$. However, we find that the accuracy to which we
can measure the average spin temperature is ultimately limited by the
accuracy to which we can measure the distribution of the covering
factor, the $N_{\rm HI}$ frequency distribution function and the
evolution of the radio source population as a function of redshift. By
improving our understanding of these distributions we will be able to
leverage the order-of-magnitude increases in sensitivity and redshift
coverage of the future SKA telescope, allowing us to measure the
evolution of the average spin temperature to much higher redshifts.

\section*{Acknowledgements} 

We thank Robert Allison, Elaine Sadler and Michael Pracy for useful
discussions, and the anonymous referee for providing comments that
helped improve this paper. JRA acknowledges support from a Bolton
Fellowship. We have made use of \texttt{Astropy}, a
community-developed core \texttt{Python} package for astronomy
(\citealt{Astropy:2013}); NASA's Astrophysics Data System
Bibliographic Services; and the VizieR catalogue access tool, operated
at CDS, Strasbourg, France.




\bibliographystyle{mnras}
\bibliography{/Users/jra/reports_thesis_papers/bibliography/james}

\bsp	
\label{lastpage}
\end{document}